\input mtexsis

\texsis
\paper

\elevenpoint
\singlespaced

\def\tableheaddoublerule{\noalign{\vglue2pt\hrule\vskip3pt\hrule\smallskip}}
\def\tableheadsinglerule{\noalign{\medskip\hrule\smallskip}}
\def\centertab#1{\hfil#1\hfil}

\def\lefttab#1{#1\hfil}
%
\def\anp#1,#2(#3){{\rm Adv.\ Nucl.\ Phys.\ }{\bf #1}, {\rm#2} {\rm(#3)}}
\def\aip#1,#2(#3){{\rm Am.\ Inst.\ Phys.\ }{\bf #1}, {\rm#2} {\rm(#3)}}
\def\aj#1,#2(#3){{\rm Astrophys.\ J.\ }{\bf #1}, {\rm#2} {\rm(#3)}}
\def\ajs#1,#2(#3){{\rm Astrophys.\ J.\ Supp.\ }{\bf #1}, {\rm#2} {\rm(#3)}}
\def\ajl#1,#2(#3){{\rm Astrophys.\ J.\ Lett.\ }{\bf #1}, {\rm#2} {\rm(#3)}}
\def\ajp#1,#2(#3){{\rm Am.\ J.\ Phys.\ }{\bf #1}, {\rm#2} {\rm(#3)}}
\def\apny#1,#2(#3){{\rm Ann.\ Phys.\ (NY)\ }{\bf #1}, {\rm#2} {\rm(#3)}}
\def\apnyB#1,#2(#3){{\rm Ann.\ Phys.\ (NY)\ }{\bf B#1}, {\rm#2} {\rm(#3)}}
\def\apD#1,#2(#3){{\rm Ann.\ Phys.\ }{\bf D#1}, {\rm#2} {\rm(#3)}}
\def\ap#1,#2(#3){{\rm Ann.\ Phys.\ }{\bf #1}, {\rm#2} {\rm(#3)}}
\def\ass#1,#2(#3){{\rm Ap.\ Space Sci.\ }{\bf #1}, {\rm#2} {\rm(#3)}}
\def\astropp#1,#2(#3)%
    {{\rm Astropart.\ Phys.\ }{\bf #1}, {\rm#2} {\rm(#3)}}
\def\aap#1,#2(#3)%
    {{\rm Astron.\ \& Astrophys.\ }{\bf #1}, {\rm#2} {\rm(#3)}}
\def\araa#1,#2(#3)%
    {{\rm Ann.\ Rev.\ Astron.\ Astrophys.\ }{\bf #1}, {\rm#2} {\rm(#3)}}
\def\arnps#1,#2(#3)%
    {{\rm Ann.\ Rev.\ Nucl.\ and Part.\ Sci.\ }{\bf #1}, {\rm#2} {\rm(#3)}}
\def\arns#1,#2(#3)%
   {{\rm Ann.\ Rev.\ Nucl.\ Sci.\ }{\bf #1}, {\rm#2} {\rm(#3)}}
\def\cpc#1,#2(#3){{\rm Comp.\ Phys.\ Comm.\ }{\bf #1}, {\rm#2} {\rm(#3)}}
\def\cjp#1,#2(#3){{\rm Can.\ J.\ Phys.\ }{\bf #1}, {\rm#2} {\rm(#3)}}
\def\cmp#1,#2(#3){{\rm Commun.\ Math.\ Phys.\ }{\bf #1}, {\rm#2} {\rm(#3)}}
\def\cnpp#1,#2(#3)%
   {{\rm Comm.\ Nucl.\ Part.\ Phys.\ }{\bf #1}, {\rm#2} {\rm(#3)}}
\def\cnppA#1,#2(#3)%
   {{\rm Comm.\ Nucl.\ Part.\ Phys.\ }{\bf A#1}, {\rm#2} {\rm(#3)}}
\def\epjC#1,#2(#3){{\rm Eur.\ Phys.\ J.\ }{\bf C#1}, {\rm#2} {\rm(#3)}}
\def\el#1,#2(#3){{\rm Europhys.\ Lett.\ }{\bf #1}, {\rm#2} {\rm(#3)}}
\def\hpa#1,#2(#3){{\rm Helv.\ Phys.\ Acta }{\bf #1}, {\rm#2} {\rm(#3)}}
\def\ieeetNS#1,#2(#3)%
    {{\rm IEEE Trans.\ }{\bf NS#1}, {\rm#2} {\rm(#3)}}
\def\IEEE #1,#2(#3)%
    {{\rm IEEE }{\bf #1}, {\rm#2} {\rm(#3)}}
\def\ijar#1,#2(#3)%
  {{\rm Int.\ J.\ of Applied Rad.\ } {\bf #1}, {\rm#2} {\rm(#3)}}
\def\ijari#1,#2(#3)%
  {{\rm Int.\ J.\ of Applied Rad.\ and Isotopes\ } {\bf #1}, {\rm#2} {\rm(#3)}}
\def\jcp#1,#2(#3){{\rm J.\ Chem.\ Phys.\ }{\bf #1}, {\rm#2} {\rm(#3)}}
\def\jgr#1,#2(#3){{\rm J.\ Geophys.\ Res.\ }{\bf #1}, {\rm#2} {\rm(#3)}}
\def\jetp#1,#2(#3){{\rm Sov.\ Phys.\ JETP\ }{\bf #1}, {\rm#2} {\rm(#3)}}
\def\jetpl#1,#2(#3)%
   {{\rm Sov.\ Phys.\ JETP Lett.\ }{\bf #1}, {\rm#2} {\rm(#3)}}
\def\jpA#1,#2(#3){{\rm J.\ Phys.\ }{\bf A#1}, {\rm#2} {\rm(#3)}}
\def\jpG#1,#2(#3){{\rm J.\ Phys.\ }{\bf G#1}, {\rm#2} {\rm(#3)}}
\def\jpamg#1,#2(#3)%
    {{\rm J.\ Phys.\ A: Math.\ and Gen.\ }{\bf #1}, {\rm#2} {\rm(#3)}}
\def\jpcrd#1,#2(#3)%
    {{\rm J.\ Phys.\ Chem.\ Ref.\ Data\ } {\bf #1}, {\rm#2} {\rm(#3)}}
\def\jpsj#1,#2(#3){{\rm J.\ Phys.\ Soc.\ Jpn.\ }{\bf G#1}, {\rm#2} {\rm(#3)}}
\def\lnc#1,#2(#3){{\rm Lett.\ Nuovo Cimento\ } {\bf #1}, {\rm#2} {\rm(#3)}}
\def\nature#1,#2(#3){{\rm Nature} {\bf #1}, {\rm#2} {\rm(#3)}}
\def\nc#1,#2(#3){{\rm Nuovo Cimento} {\bf #1}, {\rm#2} {\rm(#3)}}
\def\nim#1,#2(#3)%
   {{\rm Nucl.\ Instrum.\ Methods\ }{\bf #1}, {\rm#2} {\rm(#3)}}
\def\nimA#1,#2(#3)%
    {{\rm Nucl.\ Instrum.\ Methods\ }{\bf A#1}, {\rm#2} {\rm(#3)}}
\def\nimB#1,#2(#3)%
    {{\rm Nucl.\ Instrum.\ Methods\ }{\bf B#1}, {\rm#2} {\rm(#3)}}
\def\np#1,#2(#3){{\rm Nucl.\ Phys.\ }{\bf #1}, {\rm#2} {\rm(#3)}}
\def\mnras#1,#2(#3){{\rm ASK.\ GEORGE.\ }{\bf #1}, {\rm#2} {\rm(#3)}}
\def\medp#1,#2(#3){{\rm Med.\ Phys.\ }{\bf #1}, {\rm#2} {\rm(#3)}}
\def\mplA#1,#2(#3){{\rm Mod.\ Phys.\ Lett.\ }{\bf A#1}, {\rm#2} {\rm(#3)}}
\def\npA#1,#2(#3){{\rm Nucl.\ Phys.\ }{\bf A#1}, {\rm#2} {\rm(#3)}}
\def\npB#1,#2(#3){{\rm Nucl.\ Phys.\ }{\bf B#1}, {\rm#2} {\rm(#3)}}
\def\npBps#1,#2(#3){{\rm Nucl.\ Phys.\ (Proc.\ Supp.) }{\bf B#1},
{\rm#2} {\rm(#3)}}
\def\pasp#1,#2(#3){{\rm Pub.\ Astron.\ Soc.\ Pac.\ }{\bf #1}, {\rm#2} {\rm(#3)}}
\def\pl#1,#2(#3){{\rm Phys.\ Lett.\ }{\bf #1}, {\rm#2} {\rm(#3)}}
\def\fp#1,#2(#3){{\rm Fortsch.\ Phys.\ }{\bf #1}, {\rm#2} {\rm(#3)}}
\def\ijmpA#1,#2(#3)%
   {{\rm Int.\ J.\ Mod.\ Phys.\ }{\bf A#1}, {\rm#2} {\rm(#3)}}
\def\ijmpE#1,#2(#3)%
   {{\rm Int.\ J.\ Mod.\ Phys.\ }{\bf E#1}, {\rm#2} {\rm(#3)}}
\def\plB#1,#2(#3){{\rm Phys.\ Lett.\ }{\bf B#1}, {\rm#2} {\rm(#3)}}
\def\pnasus#1,#2(#3)%
   {{\it Proc.\ Natl.\ Acad.\ Sci.\ \rm (US)}{B#1}, {\rm#2} {\rm(#3)}}
\def\ppsA#1,#2(#3){{\rm Proc.\ Phys.\ Soc.\ }{\bf A#1}, {\rm#2} {\rm(#3)}}
\def\ppsB#1,#2(#3){{\rm Proc.\ Phys.\ Soc.\ }{\bf B#1}, {\rm#2} {\rm(#3)}}
\def\pr#1,#2(#3){{\rm Phys.\ Rev.\ }{\bf #1}, {\rm#2} {\rm(#3)}}
\def\prA#1,#2(#3){{\rm Phys.\ Rev.\ }{\bf A#1}, {\rm#2} {\rm(#3)}}
\def\prB#1,#2(#3){{\rm Phys.\ Rev.\ }{\bf B#1}, {\rm#2} {\rm(#3)}}
\def\prC#1,#2(#3){{\rm Phys.\ Rev.\ }{\bf C#1}, {\rm#2} {\rm(#3)}}
\def\prD#1,#2(#3){{\rm Phys.\ Rev.\ }{\bf D#1}, {\rm#2} {\rm(#3)}}
\def\prept#1,#2(#3){{\rm Phys.\ Reports\ } {\bf #1}, {\rm#2} {\rm(#3)}}
\def\preptC#1,#2(#3){{\rm Phys.\ Reports\ } {\bf C#1}, {\rm#2} {\rm(#3)}}
\def\prslA#1,#2(#3)%
   {{\rm Proc.\ Royal Soc.\ London }{\bf A#1}, {\rm#2} {\rm(#3)}}
\def\prl#1,#2(#3){{\rm Phys.\ Rev.\ Lett.\ }{\bf #1}, {\rm#2} {\rm(#3)}}
\def\ps#1,#2(#3){{\rm Phys.\ Scripta\ }{\bf #1}, {\rm#2} {\rm(#3)}}
\def\ptp#1,#2(#3){{\rm Prog.\ Theor.\ Phys.\ }{\bf #1}, {\rm#2} {\rm(#3)}}
\def\ppnp#1,#2(#3)%
	{{\rm Prog.\ in Part.\ Nucl.\ Phys.\ }{\bf #1}, {\rm#2} {\rm(#3)}}
\def\ptps#1,#2(#3)%
   {{\rm Prog.\ Theor.\ Phys.\ Supp.\ }{\bf #1}, {\rm#2} {\rm(#3)}}
\def\pw#1,#2(#3){{\rm Part.\ World\ }{\bf #1}, {\rm#2} {\rm(#3)}}
\def\pzetf#1,#2(#3)%
   {{\rm Pisma Zh.\ Eksp.\ Teor.\ Fiz.\ }{\bf #1}, {\rm#2} {\rm(#3)}}
\def\rgss#1,#2(#3){{\rm Revs.\ Geophysics \& Space Sci.\ }{\bf #1},
        {\rm#2} {\rm(#3)}}
\def\rmp#1,#2(#3){{\rm Rev.\ Mod.\ Phys.\ }{\bf #1}, {\rm#2} {\rm(#3)}}
\def\rnc#1,#2(#3){{\rm Riv.\ Nuovo Cimento\ } {\bf #1}, {\rm#2} {\rm(#3)}}
\def\rpp#1,#2(#3)%
    {{\rm Rept.\ on Prog.\ in Phys.\ }{\bf #1}, {\rm#2} {\rm(#3)}}
\def\science#1,#2(#3){{\rm Science\ } {\bf #1}, {\rm#2} {\rm(#3)}}
\def\sjnp#1,#2(#3)%
   {{\rm Sov.\ J.\ Nucl.\ Phys.\ }{\bf #1}, {\rm#2} {\rm(#3)}}
\def\sjpn#1,#2(#3)%
   {{\rm Sov.\ J.\ Part.\ Nucl.\ }{\bf #1}, {\rm#2} {\rm(#3)}}
\def\panp#1,#2(#3)%
   {{\rm Phys.\ Atom.\ Nucl.\ }{\bf #1}, {\rm#2} {\rm(#3)}}
\def\spu#1,#2(#3){{\rm Sov.\ Phys.\ Usp.\ }{\bf #1}, {\rm#2} {\rm(#3)}}
\def\surveyHEP#1,#2(#3)%
    {{\rm Surv.\ High Energy Physics\ } {\bf #1}, {\rm#2} {\rm(#3)}}
\def\yf#1,#2(#3){{\rm Yad.\ Fiz.\ }{\bf #1}, {\rm#2} {\rm(#3)}}
\def\zetf#1,#2(#3)%
   {{\rm Zh.\ Eksp.\ Teor.\ Fiz.\ }{\bf #1}, {\rm#2} {\rm(#3)}}
\def\zp#1,#2(#3){{\rm Z.~Phys.\ }{\bf #1}, {\rm#2} {\rm(#3)}}
\def\zpA#1,#2(#3){{\rm Z.~Phys.\ }{\bf A#1}, {\rm#2} {\rm(#3)}}
\def\zpC#1,#2(#3){{\rm Z.~Phys.\ }{\bf C#1}, {\rm#2} {\rm(#3)}}
%

\def\calrmB{\hbox\bgroup B\egroup}

\referencelist


\reference{CKMmini}
F. J. Gilman, K. Kleinknecht and B. Renk,  "CKM Quark Mixing Matrix",
to appear in the {\it Review of Particle Properties}, 2004 edition
\endreference

\reference{CPmini}
D. Kirkby and Y. Nir , "CP Violation" ,
to appear in the {\it Review of Particle Properties}, 2004 edition
\endreference

\reference{VCBmini}
M. Artuso and E. Barberio, "Determination of $|V_{\lowercase{cb}}|$",
to appear in the {\it Review of Particle Properties}, 2004 edition
\endreference


\reference{Bigi:1993ex}
I.~I.~Y.~Bigi, M.~A.~Shifman, N.~G.~Uraltsev and A.~I.~Vainshtein,
Int.\ J.\ Mod.\ Phys.\ A {\bf 9}, 2467 (1994)
[arXiv:hep-ph/9312359]
\endreference

\reference{Neubert:1996qg}
M.~Neubert,
Int.\ J.\ Mod.\ Phys.\ A {\bf 11}, 4173 (1996)
[arXiv:hep-ph/9604412]
\endreference

\reference{Bigi:1997dn}
I.~I.~Bigi, M.~A.~Shifman and N.~Uraltsev,
Ann.\ Rev.\ Nucl.\ Part.\ Sci.\  {\bf 47}, 591 (1997)
[arXiv:hep-ph/9703290]
\endreference

\reference{Hoang:1998ng}
A.~H.~Hoang, Z.~Ligeti and A.~V.~Manohar,
Phys.\ Rev.\ Lett.\  {\bf 82}, 277 (1999)
[arXiv:hep-ph/9809423]
\endreference

\reference{Bigihepph9907270} {I.I. Bigi,
UND-HEP-BIG-99-05, {\tt hep-ph/9907270\rm }}
\endreference

\reference{Zoltanhepph9908432} {Z. Ligeti,
FERMILAB-Conf-99/213-T, {\tt hep-ph/9908432\rm }}
\endreference

\reference{Ritbergen:1999}
T.~van Ritbergen, Phys.\ Lett.\ B {\bf
  454}, 353 (1999) [hep-ph/9903226]
\endreference
 
\reference{Battaglia:2003in}
M.~Battaglia {\it et al.},
arXiv:hep-ph/0304132
\endreference

\reference{bb:DurhamCKMProceedings}
Proceedings of the Second Workshop on the CKM Unitarity Triangle, Durham, 2003,
edited by P. Ball, J. Flynn, P. Kluit and A. Stocchi, eConf C0304052 (2003)
\endreference

\reference{hfworkinggroup}
The LEP VUB Working Group, Note LEPVUB-01/01
\endreference

\reference{bb:luke_elkhadra_bmass_annrev}
A.X.~El-Khadra and M. Luke,
Ann. Rev. Nucl. Sci. {\bf 52}, 201 (2002)
\endreference


\reference{Bigi:2001ys}
I.~I.~Y.~Bigi and N.~Uraltsev,
Int.\ J.\ Mod.\ Phys.\ A {\bf 16}, 5201 (2001)
[arXiv:hep-ph/0106346]
\endreference

\reference{bb:luke_ckm}
M.~Luke,
eConf {\bf C0304052}, WG107 (2003)
[arXiv:hep-ph/0307378]
\endreference

\reference{Ligeti:2003hp}
Z.~Ligeti,
arXiv:hep-ph/0309219
\endreference

\reference{bb:cleo_endpoint}
A.~Bornheim {\it et al.}  [CLEO Collaboration],
Phys.\ Rev.\ Lett.\  {\bf 88}, 231803 (2002)
[arXiv:hep-ex/0202019]
\endreference

\reference{bb:babar_endpoint}
B.~Aubert {\it et al.}  [BABAR Collaboration],
arXiv:hep-ex/0207081
\endreference

\reference{bb:belle_endpoint}
K.~Abe {\it et al.} [BELLE Collaboration],
BELLE-CONF-0325, 2003
\endreference

\reference{bb:cleo_orig_endpoint}{R. Fulton \etal [CLEO Collab.],
\prl64,16,(1990) and  J. Bartelt \etal, 
\prl71,4111(1993)}
\endreference

\reference{bb:argus_orig_endpoint}{H. Albrecht \etal [ARGUS Collab.]
\plB234,409(1990) and 
\plB255,297(1991)}
\endreference

\reference{bb:babar_mx}
B.~Aubert {\it et al.}  [BABAR Collaboration],
arXiv:hep-ex/0307062
\endreference

\reference{bb:belle_mx}
C.~Schwanda ({\it for the BELLE Collaboration})
to appear in the ``Proceedings of the International Europhysics 
Conference on High Energy Physics -- EPS 2003'',
Aachen, Germany, July, 2003
\endreference

\reference{bkp90}{V. Barger, C.S. Kim, and R.J.N. Phillips,
\plB251, 629 (1990)}
\endreference

\reference{ligetifalkwise}
A.F.~Falk, Z.~Ligeti and M.B.~Wise,
\plB406, 225 (1997)
\endreference

\reference{bdu98}
{I.I. Bigi, R.D. Dikeman, and N. Uraltsev, 
Eur.\ Phys.\ J.\ {\bf C4}, 453 (1998)}
\endreference

\reference{NeubertDeFazio}
F.~De Fazio and M.~Neubert,
JHEP {\bf 9906}, 017 (1999)
\endreference

\reference{bb:delphi_mx}
{P.~Abreu {\it et al.} [DELPHI Collaboration], 
\plB478, 14 (2000)}
\endreference

\reference{neubertsf}
M.~Neubert,
Phys.\ Rev.\ D {\bf 49}, 3392 (1994)
[arXiv:hep-ph/9311325]
\endreference

\reference{bigiEndpoint}{I. Bigi, M. Shifman, N. Uraltsev and
A. Vainshtein, Int.\ J.\ Mod.\ Phys.\  {\bf A9}, 2467 (1994)}
\endreference

\reference{neubertEndpoint}{M. Neubert
\prD49, 4623 (1994)}
\endreference

\reference{Aglietti:2000te}
U.~Aglietti and G.~Ricciardi,
Nucl.\ Phys.\ B {\bf 587}, 363 (2000)
[arXiv:hep-ph/0003146]
\endreference

\reference{Dikeman96}
R.~D.~Dikeman, M.~A.~Shifman and N.~G.~Uraltsev,
Int.\ J.\ Mod.\ Phys.\ A {\bf 11}, 571 (1996)
[arXiv:hep-ph/9505397]
\endreference

\reference{Leibovich:2000kv}
A.~K.~Leibovich,
in {\it Proc. of the 5th International Symposium on Radiative Corrections (RADCOR 2000) } ed. Howard E. Haber,
[arXiv:hep-ph/0011181]
\endreference

\reference{neubertNote}
M.~Neubert,
\plB513, 88 (2001)
\endreference

\reference{bb:rothstein_endpoint}
A.~K.~Leibovich, I.~Low and I.~Z.~Rothstein,
Phys.\ Rev.\ D {\bf 61}, 053006 (2000)
[arXiv:hep-ph/9909404]
\endreference

\reference{bb:Rothstein_MX}
A.~K.~Leibovich, I.~Low and I.~Z.~Rothstein,
Phys.\ Lett.\ B {\bf 486}, 86 (2000)
[arXiv:hep-ph/0005124]
\endreference

\reference{bb:neubert_integ}
M.~Neubert,
Phys.\ Lett.\ B {\bf 513}, 88 (2001)
[arXiv:hep-ph/0104280]
\endreference

\reference{bb:llr_comment}
A.~K.~Leibovich, I.~Low and I.~Z.~Rothstein,
Phys.\ Lett.\ B {\bf 513}, 83 (2001)
[arXiv:hep-ph/0105066]
\endreference

\reference{bb:bigi_weighted_integrals}
I.~Bigi and N.~Uraltsev,
Int.\ J.\ Mod.\ Phys.\ A {\bf 17}, 4709 (2002)
[arXiv:hep-ph/0202175]
\endreference

\reference{bb:hepph0107074}
C.~W.~Bauer, Z.~Ligeti and M.~E.~Luke,
Phys.\ Rev.\ D {\bf 64}, 113004 (2001)
[arXiv:hep-ph/0107074]
\endreference

\reference{bb:bll1}
C.~W.~Bauer, Z.~Ligeti and M.~E.~Luke,
Phys.\ Lett.\ B {\bf 479}, 395 (2000)
[arXiv:hep-ph/0002161]
\endreference

\reference{bb:original_neubert_mccubed}
M.~Neubert,
JHEP {\bf 0007}, 022 (2000)
[arXiv:hep-ph/0006068]
\endreference

\reference{bb:neubert_mccubed}
M.~Neubert and T.~Becher,
Phys.\ Lett.\ B {\bf 535}, 127 (2002)
[arXiv:hep-ph/0105217]
\endreference

\reference{bb:belle_mxq2}
H.~Kakuno {\it et al.}  [BELLE Collaboration],
arXiv:hep-ex/0311048
\endreference

\reference{bigi94}
I.I.~Bigi, M.A.~Shifman, N.~Uraltsev and A.I.~Vainshtein,
\plB328, 431 (1994)
\endreference

\reference{kagan98}
A.L.~Kagan and M.~Neubert,
\epjC7, 5 (1999)
\endreference

\reference{bb:aMatthiasReference}
M.~Neubert, private communications and CLNS-04/1858 (in preparation)
\endreference

\reference{bb:hepph0312109}
C.~W.~Bauer and A.~V.~Manohar,
arXiv:hep-ph/0312109
\endreference

\reference{bb:hfag_vub}
http://www.slac.stanford.edu/xorg/hfag/semi/summer03-lp/summer03.shtml
\endreference

\reference{bb:gibbonsbeauty}{L. Gibbons,
to appear in "Proceedings of the 9th International Conference on 
B-Physics at Hadron Machines - BEAUTY 2003", Carnegie-Mellon University, 2003,
arXiv:hep-ex/0402009.
L. Gibbons, D. Hennessy and E. Thorndike, in preparation}
\endreference

\reference{alephvub}
{R.~Barate {\it et al.} [ALEPH Collab.], 
Eur.\ Phys.\ J.\ {\bf C6}, 555 (1999)}
\endreference

\reference{l3vub}
{M.~Acciarri {\it et al.} [L3 Collab.],  
\plB436, 174  (1998)}
\endreference

\reference{opalvub}
G.~Abbiendi {\it et al.} [OPAL Collab.], 
Eur.\ Phys.\ J.\ {\bf C21} 399 (2001)
\endreference

\reference{cleo_triple_vub}
A.~Bornheim {\it et al.} [CLEO Collaboration],
CLEO-CONF-02-08, 2002
\endreference

\reference{LLW_SSF}
A.~K.~Leibovich, Z.~Ligeti and M.~B.~Wise,
Phys.\ Lett.\ B {\bf 539}, 242 (2002)
[arXiv:hep-ph/0205148]
\endreference

\reference{BLT_SSF2}
C.~W.~Bauer, M.~Luke and T.~Mannel,
Phys.\ Lett.\ B {\bf 543}, 261 (2002)
[arXiv:hep-ph/0205150]
\endreference

\reference{N_SSF}
M.~Neubert,
Phys.\ Lett.\ B {\bf 543}, 269 (2002)
[arXiv:hep-ph/0207002]
\endreference

\reference{BLT_SSF1}
C.~W.~Bauer, M.~E.~Luke and T.~Mannel,
Phys.\ Rev.\ D {\bf 68}, 094001 (2003)
[arXiv:hep-ph/0102089]
\endreference

\reference{bigi_wa}
 I.I.~Bigi and N.G.~Uraltsev, 
  Nucl.\ Phys.\ B {\bf 423} (1994) 33 [hep-ph/9310285]
\endreference
  
\reference{voloshin_wa}
M.~B.~Voloshin,
Phys.\ Lett.\ B {\bf 515}, 74 (2001)
[arXiv:hep-ph/0106040]
\endreference


\reference{Gilman:1989uy}
F.~J.~Gilman and R.~L.~Singleton,
\prD41,142(1990)
\endreference

%
%

\reference{Abada:1993dh}
A.~Abada {\it et al.},
Nucl.\ Phys.\ B {\bf 416} (1994) 675
[arXiv:hep-lat/9308007]
\endreference

\reference{Allton:1994ui}
C.~R.~Allton {\it et al.}  [APE Collab.],
Phys.\ Lett.\ B {\bf 345}, 513 (1995)
[arXiv:hep-lat/9411011]
\endreference

\reference{DelDebbio:1997kr}
L.~Del Debbio, J.~M.~Flynn, L.~Lellouch and J.~Nieves  [UKQCD Collab.],
Phys.\ Lett.\ B {\bf 416}, 392 (1998)
[arXiv:hep-lat/9708008]
\endreference

\reference{Hashimoto:1997sr}
S.~Hashimoto, K.~I.~Ishikawa, H.~Matsufuru, T.~Onogi and N.~Yamada,
Phys.\ Rev.\ D {\bf 58}, 014502 (1998)
[arXiv:hep-lat/9711031]
\endreference

\reference{Ryan:1998tj}
S.~Ryan, A.~El-Khadra, S.~Hashimoto, A.~Kronfeld, P.~Mackenzie and
J.~Simone,
Nucl.\ Phys.\ Proc.\ Suppl.\  {\bf 73}, 390 (1999)
[arXiv:hep-lat/9810041]
\endreference

\reference{Ryan:1999kx}
S.~M.~Ryan, A.~X.~El-Khadra, A.~S.~Kronfeld, P.~B.~Mackenzie and
J.~N.~Simone,
Nucl.\ Phys.\ Proc.\ Suppl.\  {\bf 83}, 328 (2000)
[arXiv:hep-lat/9910010]
\endreference

\reference{Lellouch:1999dz}
L.~Lellouch,
[arXiv:hep-ph/9912353]
\endreference

\reference{Bowler:1999xn}
K.~C.~Bowler {\it et al.}  [UKQCD Collab.],
Phys.\ Lett.\ B {\bf 486}, 111 (2000)
[arXiv:hep-lat/9911011]
\endreference

\reference{Becirevic:1999kt}
D.~Becirevic and A.~B.~Kaidalov,
Phys.\ Lett.\ B {\bf 478}, 417 (2000)
[arXiv:hep-ph/9904490]
\endreference

\reference{Aoki:2000by}
S.~Aoki {\it et al.}  [JLQCD Collab.],
Nucl.\ Phys.\ Proc.\ Suppl.\  {\bf 94}, 329 (2001)
[arXiv:hep-lat/0011008]
\endreference

\reference{Abada:2000ty}
A.~Abada, D.~Becirevic, P.~Boucaud, J.~P.~Leroy, V.~Lubicz and
F.~Mescia,
Nucl.\ Phys.\ B {\bf 619}, 565 (2001)
[arXiv:hep-lat/0011065]
\endreference

\reference{El-Khadra:2001rv}
A.~X.~El-Khadra, A.~S.~Kronfeld, P.~B.~Mackenzie, S.~M.~Ryan and
J.~N.~Simone,
Phys.\ Rev.\ D {\bf 64}, 014502 (2001)
[arXiv:hep-ph/0101023]
\endreference

\reference{Aoki:2001rd}
S.~Aoki {\it et al.}  [JLQCD Collab.],
Phys.\ Rev.\ D {\bf 64}, 114505 (2001)
[arXiv:hep-lat/0106024]
\endreference

%
%

\reference{Ball:1997rj}
P.~Ball and V.~M.~Braun,
Phys.\ Rev.\ D {\bf 55}, 5561 (1997)
[arXiv:hep-ph/9701238]
\endreference

\reference{Ball:1998kk}
P.~Ball and V.~M.~Braun,
Phys.\ Rev.\ D {\bf 58}, 094016 (1998)
[arXiv:hep-ph/9805422]
\endreference

\reference{Khodjamirian:1997ub}
A.~Khodjamirian, R.~Ruckl, S.~Weinzierl and O.~I.~Yakovlev,
Phys.\ Lett.\ B {\bf 410}, 275 (1997)
[arXiv:hep-ph/9706303]
\endreference

\reference{Khodjamirian:2000ds}
A.~Khodjamirian, R.~Ruckl, S.~Weinzierl, C.~W.~Winhart and
O.~I.~Yakovlev,
Phys.\ Rev.\ D {\bf 62}, 114002 (2000)
[arXiv:hep-ph/0001297]
\endreference

\reference{Bakulev:2000fb}
A.~P.~Bakulev, S.~V.~Mikhailov and R.~Ruskov,
[arXiv:hep-ph/0006216]
\endreference

\reference{Huang:2000hs}
T.~Huang, Z.~Li and X.~Wu,
[arXiv:hep-ph/0011161]
\endreference

\reference{Wang:2001mi}
W.~Y.~Wang and Y.~L.~Wu,
Phys.\ Lett.\ B {\bf 515}, 57 (2001)
[arXiv:hep-ph/0105154]
\endreference

\reference{Wang:2001bh}
W.~Y.~Wang and Y.~L.~Wu,
Phys.\ Lett.\ B {\bf 519}, 219 (2001)
[arXiv:hep-ph/0106208]
\endreference

\reference{Ball:2001fp}
P.~Ball and R.~Zwicky,
JHEP {\bf 0110}, 019 (2001)
[arXiv:hep-ph/0110115]
\endreference

%
%

\reference{Wirbel:1985ji}
M.~Wirbel, B.~Stech and M.~Bauer,
Z.\ Phys.\ C {\bf 29}, 637 (1985)
\endreference

\reference{Korner:1987kd}
J.~G.~Korner and G.~A.~Schuler,
Z.\ Phys.\ C {\bf 38}, 511 (1988)
[Erratum-ibid.\ C {\bf 41}, 690 (1988)]
\endreference

\reference{Isgur:gb}
N.~Isgur, D.~Scora, B.~Grinstein and M.~B.~Wise,
Phys.\ Rev.\ D {\bf 39}, 799 (1989)
\endreference

\reference{Scora:1995ty}
D.~Scora and N.~Isgur,
Phys.\ Rev.\ D {\bf 52}, 2783 (1995)
[arXiv:hep-ph/9503486]
\endreference

\reference{Melikhov:1995xz}
D.~Melikhov,
Phys.\ Rev.\ D {\bf 53}, 2460 (1996)
[arXiv:hep-ph/9509268]
\endreference

\reference{Beyer:1998ka}
M.~Beyer and D.~Melikhov,
Phys.\ Lett.\ B {\bf 436}, 344 (1998)
[arXiv:hep-ph/9807223]
\endreference

\reference{Faustov:1995bf}
R.~N.~Faustov, V.~O.~Galkin and A.~Y.~Mishurov,
Phys.\ Rev.\ D {\bf 53}, 6302 (1996)
[arXiv:hep-ph/9508262]
\endreference

\reference{Demchuk:1997uz}
N.~B.~Demchuk, P.~Y.~Kulikov, I.~M.~Narodetsky and P.~J.~O'Donnell,
Phys.\ Atom.\ Nucl.\  {\bf 60}, 1292 (1997)
[Yad.\ Fiz.\  {\bf 60N8}, 1429 (1997)]
[arXiv:hep-ph/9701388]
\endreference

\reference{Grach:1996nz}
I.~L.~Grach, I.~M.~Narodetsky and S.~Simula,
Phys.\ Lett.\ B {\bf 385}, 317 (1996)
[arXiv:hep-ph/9605349]
\endreference

\reference{:2000ae}
Riazuddin, T.~A.~Al-Aithan and A.~H.~S.~Gilani,
Int.\ J.\ Mod.\ Phys.\ A {\bf 17}, 4927 (2002)
[arXiv:hep-ph/0007164]
\endreference

\reference{Melikhov:2000yu}
D.~Melikhov and B.~Stech,
Phys.\ Rev.\ D {\bf 62}, 014006 (2000)
[arXiv:hep-ph/0001113]
\endreference

\reference{Feldmann:1999sm}
T.~Feldmann and P.~Kroll,
Eur.\ Phys.\ J.\ C {\bf 12}, 99 (2000)
[arXiv:hep-ph/9905343]
\endreference

\reference{Flynn:2000gd}
J.~M.~Flynn and J.~Nieves,
Phys.\ Lett.\ B {\bf 505}, 82 (2001)
[arXiv:hep-ph/0007263]
\endreference

\reference{Beneke:2000wa}
M.~Beneke and T.~Feldmann,
Nucl.\ Phys.\ B {\bf 592}, 3 (2001)
[arXiv:hep-ph/0008255]
\endreference

\reference{Choi:1999nu}
H.~M.~Choi and C.~R.~Ji,
Phys.\ Lett.\ B {\bf 460}, 461 (1999)
[arXiv:hep-ph/9903496]
\endreference

%
%

\reference{Kurimoto:2001zj}
T.~Kurimoto, H.~n.~Li and A.~I.~Sanda,
Phys.\ Rev.\ D {\bf 65}, 014007 (2002)
[arXiv:hep-ph/0105003]
\endreference

%
%

\reference{Ligeti:1995yz}
Z.~Ligeti and M.~B.~Wise,
Phys.\ Rev.\ D {\bf 53}, 4937 (1996)
[arXiv:hep-ph/9512225]
\endreference

\reference{Aitala:1997cm}
E.~M.~Aitala {\it et al.}  [E791 Collab.],
Phys.\ Rev.\ Lett.\  {\bf 80}, 1393 (1998)
[arXiv:hep-ph/9710216]
\endreference

%
%

\reference{Burdman:1996kr}
G.~Burdman and J.~Kambor,
Phys.\ Rev.\ D {\bf 55}, 2817 (1997)
[arXiv:hep-ph/9602353]
\endreference

\reference{Lellouch:1995yv}
L.~Lellouch,
Nucl.\ Phys.\ B {\bf 479}, 353 (1996)
[arXiv:hep-ph/9509358]
\endreference

\reference{Mannel:1998kp}
T.~Mannel and B.~Postler,
Nucl.\ Phys.\ B {\bf 535}, 372 (1998)
[arXiv:hep-ph/9805425]
\endreference

\reference{bb:ph0310139}
B.~O.~Lange,
arXiv:hep-ph/0310139
\endreference

\reference{bb:ph0308249}
P.~Ball,
arXiv:hep-ph/0308249
\endreference

\reference{bb:ph0311335}
M.~Beneke and T.~Feldmann,
arXiv:hep-ph/0311335
\endreference

\reference{bb:ph0311345}
B.~O.~Lange and M.~Neubert,
arXiv:hep-ph/0311345
\endreference

\reference{bb:lat0304004}
C.~T.~H.~Davies {\it et al.}  [HPQCD Collaboration],
arXiv:hep-lat/0304004
\endreference

\reference{bb:lat0309055}
C.~Bernard {\it et al.}  [MILC Collaboration],
arXiv:hep-lat/0309055\endreference

\reference{bb:lat0309107}
M.~Okamoto {\it et al.},
arXiv:hep-lat/0309107
\endreference

\reference{cleo-pilnu}
S.~B.~Athar {\it et al.}  [CLEO Collaboration],
Phys.\ Rev.\ D {\bf 68}, 072003 (2003)
[arXiv:hep-ex/0304019]
\endreference

\reference{cleo-rholnu}{B.H. Behrens \etal [CLEO Collab.],
\prD61, 052001 (2000)}
\endreference

\reference{babar-rholnu}
B.~Aubert {\it et al.}  [BABAR Collaboration],
Phys.\ Rev.\ Lett.\  {\bf 90}, 181801 (2003)
[arXiv:hep-ex/0301001]
\endreference

\reference{bb:DiazCruz}
J.~L.~Diaz-Cruz, G.~Lopez Castro, and J.~H.~Munoz,
Phys.\ Rev.\ D {\bf 54}, 2388 (1996)
[arXiv:hep-ph/9605344]
\endreference

\reference{bb:gibbons_ckm}
L.~Gibbons,
eConf {\bf C0304052}, WG105 (2003)
[arXiv:hep-ex/0307065]
\endreference

\reference{bb:belle_omega}
K.~Abe {\it et al.}  [Belle Collaboration],
arXiv:hep-ex/0307075
\endreference

\reference{bb:lepage_moving_B}
K.~M.~Foley and G.~P.~Lepage,
Nucl.\ Phys.\ Proc.\ Suppl.\  {\bf 119}, 635 (2003)
[arXiv:hep-lat/0209135]
\endreference

\endreferencelist

{ 
\baselineskip = 0pt
\hfil{\twelvepoint CLNS 04/1864}\break

\hfil{\twelvepoint LBNL-54491}\break
\vskip 12 pt
}

\titlepage
\title{Determination of $|V_{\lowercase{ub}}|$}

\author
M.~Battaglia
Dept. of Physics, University of California at Berkeley and 
Lawrence Berkeley National Laboratory 
Berkeley, CA.
\endauthor
\and
\author
L.~Gibbons
Dept. of Physics, Cornell University
 Ithaca, NY
 \endauthor
 
 \abstract
The element $V_{ub}$ of the CKM mixing matrix is the smallest of the quark couplings 
and is crucial in understanding  CP violation in the $B$ system.  In this review, we 
discuss the present status of the determination of its magnitude, which  involves a significant effort 
both experimentally and  theoretically. Decisive progress has been achieved in 
recent years, thanks in part to the large data sets that have become available at 
the $B$ factories. Based on these data we propose an average $|V_{ub}|$ value of
$(3.67 \pm 0.47) \times 10^{-3}$, which has an uncertainty of 13~\%. With the 
experimental and  theoretical progress expected in the next few years 
a determination to an accuracy of 10~\%, or better,  seems feasible.
\endabstract
 
 \toappear{Review of Particle Properties}
 
\endtitlepage
 
\def\Vub{|V_{ub}|}
\def\Vcb{|V_{cb}|}
\def\btoulnu{b\rightarrow u \ell \bar \nu}
\def\btoclnu{b\rightarrow c \ell \bar \nu}
\medskip
\noindent

The precise determination of the magnitude of $V_{ub}$ with a robust, 
well-understood uncertainty remains one of the
key goals of the heavy flavor physics programs, both experimentally and theoretically.
Because $\Vub$, the smallest element in the 
CKM mixing matrix,  provides a bound on the upper vertex of one of the triangles 
representing the unitarity property of the CKM matrix, it plays a 
crucial role in the examination of the unitarity constraints and the fundamental 
questions on which the constraints can bear (see the minireviews on the CKM matrix\cite{CKMmini}
and on $CP$--violation\cite{CPmini}
for details). Investigation of these issues require 
measurements that are precise and that have 
well--understood uncertainties.  

The charmless semi-leptonic (s.l.) 
decay channel $b \rightarrow u \ell \bar \nu$
provides the cleanest path for the determination 
of $\Vub$.  However, the theory for the heavy--to-light $b\to u$ transition cannot 
be as well constrained as  that for 
the heavy--to-heavy $b\to c$ transition used in the determination of $|V_{cb}|$ 
(see the $|V_{cb}|$ minireview\cite{VCBmini}).
The extraction of $\Vub$ and the interplay between experimental measurements and their 
theoretical interpretation are further complicated by the large
background from $b\to c\ell \bar{\nu}$ decay, which has a rate about 
60 times higher than that for charmless s.l.\ decay. 
Measurements based both on exclusive decay channels 
and on inclusive techniques have been, and are being, pursued. 

The last several years have seen significant developments in both the theoretical
framework and the experimental techniques used in the study of $\btoulnu$.
The inclusive theory has progressed significantly in categorization 
of the corrections to the base theory still needed, in their relative importance 
in different regions of phase space, and in the determination of some of them. 
Recent work on exclusive processes bolsters confidence in 
the current uncertainties for the form factor calculations needed to extract $\Vub$. 
Experimentally, we have new inclusive and exclusive measurements that minimize 
dependence on detailed modeling of the signal process to separate signal 
from the $\btoclnu$ background, have well-defined sensitivities in particular regions of 
phase space and have improved signal-to-background ratios.
These improvements provide us with a first opportunity to develop a method 
for a robust determination of $\Vub$ with complete error estimates, including constraints on
hitherto unquantified contributions.  
We review the current determinations of $\Vub$, focusing primarily on these recent
developments. An average of the inclusive information from all regions of phase space
remains, unfortunately, beyond our reach because of the potentially sizable corrections
for which we lack estimates.  Rather, we combine the inclusive results to 
obtain a central value and, in particular, a more complete evaluation of the uncertainty
than has been possible in the past.   

\section{ Inclusive measurements of $\btoulnu$}

Theoretically, issues regarding the calculation of the total semileptonic 
partial width $\Gamma(B\to X_u\ell\nu)$ via the operator product expansion
(OPE) are well-understood\refrange{Bigi:1993ex}{Zoltanhepph9908432}.   
The OPE is both a nonperturbative power series in $1/m_b$ and
a perturbative expansion in $\alpha_s$.
At order $1/m_b^2$, it predicts
$$
\eqalign{
\Gamma&(B\to  X_u \ell\nu)= {G_F^2 |V_{ub}|^2\over 192\pi^3}\, m_b^5 \cr
& \quad \times\left[1-
\frac{9\lambda_2 - \lambda_1}{2m_b^2} +\ldots - {\cal O}(\frac{\alpha_s}{\pi})\right]},
\EQN{eq:parton}
$$
where $\lambda_2$ parameterizes the hyperfine interaction between the heavy
quark and the light degrees of freedom and $\lambda_1$ is related to the Fermi momentum
of the heavy quark. The perturbative corrections are known
to order $\alpha_s^2$\cite{Ritbergen:1999}.  The OPE is alternatively
expressed in terms of the nonperturbative parameters $\mu_\pi^2$ and $\mu_G^2$,
which are closely related to $\lambda_1$
and $\lambda_2$, respectively, but differ significantly in their infrared treatment.
Within the OPE, the importance of a proper field theoretic treatment of the parameters
is paramount both for the total rate and for the restricted phase space studies
discussed below.  The treatment of the quark mass and its
associated uncertainties are particularly important given the strong mass dependance 
of the width.  Such considerations have led to useful definitions like the
kinematic mass, which are discussed in detail in \Ref{Battaglia:2003in} and \Ref{bb:DurhamCKMProceedings}.

The error induced by uncertainties in the nonperturbative
 parameters $\lambda_{1,2}$ is relatively small, and an evaluation\cite{hfworkinggroup}
 by the LEP VUB working group yielded
$$ 
\eqalign{
\Vub &= 0.00445
\left( 
\frac{{\calrmB}(b\to u\ell\bar{\nu})}{0.002} \frac{1.55{\rm ps}}{\tau_b}
\right)^{1/2} \cr
& \quad \times \left(1\pm 0.020_{\rm OPE} \pm0.052_{m_b}\right)
}
\EQN{eq:rate}
$$
The quoted uncertainty is dominated by the uncertainty in the $b$ quark mass,
for which $m^{1S}_b (1GeV ) = 4.58\pm0.09$ GeV was assumed.  The value and
uncertainty are in good agreement with a recent survey\cite{bb:luke_elkhadra_bmass_annrev}. 
No weak annihilation uncertainties (discussed below) are included in the 
quoted $OPE$ error. 
Use of the quark-level OPE for prediction of moments of the true inclusive spectra 
has generated concern regarding potential violation of the underlying assumption
of global quark-hadron duality.
This concern has been confronted by theoretical wisdom\cite{Bigi:2001ys} 
supporting global duality for the inclusive $\btoulnu$ transition, and new data both support 
this assumption and allow placement of quantitative limits on the violation. 
In particular, the exclusive and inclusive extractions of $\Vcb$ 
agree to $(0.8 \pm 1.6) \times 10^{-3}$\cite{VCBmini} .
Taking the uncertainty as an upper bound on the global duality violation 
for $\Vub$ shows that it should not exceed  $\simeq$4\%.  Using this as
bound as an uncertainty estimate for duality effects in the partial width
prediction brings the total
uncertainty on $\Vub$ to $6.8\%$. The agreement of the 
OPE parameters extracted using moments  of different distributions in s.l.\ decays 
further supports  the small scale for duality  violation effects.  

While theoretically the total inclusive rate would allow determination of $\Vub$ to
better than 10\%, experimentally the much more copious $\btoclnu$ process makes
a measurement over the full phase space unrealizable.  To overcome this
background, inclusive $\btoulnu$ measurements utilize restricted regions of phase
space in which the $\btoclnu$ process is  kinematically highly suppressed.
The background is forbidden in the regions of large charged lepton energy $E_\ell > (M^2_B - M^2_D)/(2 M_B)$ (the endpoint), 
low hadronic mass $M_X < M_D$ and large dilepton mass  $q^2 > (M_B - M_D)^2$.
Extraction of $\Vub$ from such a measurement requires knowledge of the fraction of the 
total $\btoulnu$ rate that lies within the utilized region of phase space, which complicates
the theoretical issues and uncertainty considerably.  \Ref{bb:luke_ckm} and
\Ref{Ligeti:2003hp} discuss the issues in detail.



CLEO\cite{bb:cleo_endpoint}, BaBar\cite{bb:babar_endpoint} and BELLE\cite{bb:belle_endpoint}
have all presented recent measurements of the $\btoulnu$ rate near the endpoint.  The
results, which are for integrated ranges in the $\Upsilon(4S)$ rest frame,
are summarized in 
\Tbl{tab:endpointRates}.   
Experimentally, these measurements must contend with a large background from
continuum $e^+e^-$ annihilation processes.  Suppression of these backgrounds
introduces significant efficiency variation with the $q^2$ of the decay,
which introduces model dependence.  Greater awareness of this issue has resulted
in more sophisticated suppression methods in these  recent
measurements and thus in over a factor of three reduction in the
model dependence of the measured rates
relative to earlier measurements\cite{bb:cleo_orig_endpoint}\cite{bb:argus_orig_endpoint}.
Future measurements, either using fully-reconstructed
$B$ tag samples that would remove the problem or a modest $q^2$ binning, would
essentially eliminate the remaining model dependence.


{
\midtable{tab:endpointRates}
\Caption
Partial branching fractions for $\btoulnu$ within the charged lepton momentum range
($\Upsilon(4S)$ frame) from  $2.6$ GeV/$c$ down to the indicated minimum.
The estimated fraction $f_E$ of the total
$\btoulnu$ rate expected to lie in that range is also given.  The dagger ($\dagger$) indicates the
quantity that received the QED radiative correction appropriate to the indicated mix of electrons
and muons, which has not always been treated self-consistently
in the literature. 
\endCaption
\vglue -18pt
\centerline{
\vbox{
\offinterlineskip
\halign{\strut\hfil #\hfil &\tabskip=.5em plus 1em minus 0.5em
\centertab{#}& \centertab{#} & \lefttab{#}\tabskip=0pt\cr
\tableheaddoublerule
$p_\ell^{{\rm min}}$ &	$\Delta {\cal B}_u(p)$ & $f_E$ & \cr
(GeV/$c$) & ($10^{-4}$) & & \cr
\tableheadsinglerule
 2.0	& $4.22 \pm 0.33 \pm 1.78$   & $\dagger 0.266 \pm 0.041 \pm 0.024$   & CLEO ($e,\mu$) \cr
 2.1	& $3.28 \pm 0.23 \pm 0.73$   & $\dagger 0.198 \pm 0.035 \pm 0.020$   & CLEO ($e,\mu$)\cr
 2.2	&  $2.30 \pm 0.15 \pm 0.35$   & $\dagger 0.130 \pm 0.024 \pm 0.015$   & CLEO ($e,\mu$)\cr
 2.3	& $1.43 \pm 0.10 \pm 0.13$   & $\dagger0.074 \pm 0.014 \pm 0.009$   & CLEO ($e,\mu$)\cr
    		&  $\dagger 1.52 \pm 0.14 \pm 0.14$   & $ 0.078 \pm 0.015 \pm 0.009$   & BaBar ($e$)\cr
  		&  $1.19 \pm 0.11 \pm 0.10$   & $\dagger 0.072 \pm 0.014 \pm 0.008$   & BELLE ($e$)\cr
 2.4	& $0.64 \pm 0.07 \pm 0.05$   & $\dagger 0.037 \pm 0.007 \pm 0.003$   & CLEO  ($e,\mu$)\cr
\tableheaddoublerule
} 
} 
} 
\endtable
}

BaBar\cite{bb:babar_mx} and BELLE\cite{bb:belle_mx} have presented
new analyses of the low $M_X$ 
region\refrange{bkp90}{NeubertDeFazio}. They also utilize a moderate 
(only $\sim10\%$ loss) $p_\ell>1.0$ GeV$/c$ requirement.  
This technique was pioneered by Delphi at LEP\cite{bb:delphi_mx}, but 
the achievable resolution and signal-to-background ratio were lower compared to 
those obtained at the $B$ factories. 
Because of experimental resolution on $M_X$, the $\btoclnu$ background
smears below its theoretical lower limit of $M_X=M_D$, so experiments
must impose more stringent $M_X$ requirements which are theoretically more problematic.
The BaBar and BELLE analyses are based on ``$B$ tag'' samples of
fully reconstructed hadronic decays and $D^{(*)}\ell\nu$ decays, respectively.
In both cases, $M_X$ is calculated directly from the particles remaining
after removal of  tag and lepton contributions.  
The BaBar analysis, in particular, reveals a beautiful $\btoulnu$ signal with an 
unsurpassed signal to background ratio of about 2:1 in the region $M_X<1.55$ GeV$/c$,
which rivals that of current exclusive analyses.  This analysis
demonstrates the anticipated power of a large fully reconstructed sample,
both in the signal to background levels and in
the excellent resolution that can be achieved.
The efficiency versus $M_X$
 appears reasonably uniform, and the signal yield fitting procedure
minimizes the dependence of the extracted rate on the modeling of the detailed $\btoulnu$
dynamics. Both allow for improved determination of $\Vub$ as theory advances.   

Determination of the fraction of the $\btoulnu$ rate in the $p_\ell$ endpoint or the
 low $M_X$ region requires resummation of the OPE to all orders in 
 ${E_X\Lambda_{QCD}}/{m_x^2}$\refrange{neubertsf}{Dikeman96}.
 The resummation results, at leading-twist order, in a nonperturbative shape
 function 
 $f(k_+)$, where
 $k_+ = k^0 + k_\|$ and $k^\mu = p_b^\mu - m_b v^\mu$ is the
 residual $b$ quark momentum after the ``mechanical'' portion of
   momentum is subtracted off.  Spatial components 
 $k_\|$ and $k_\perp$ are defined relative to the
  $m_b v^\mu - q^\mu$ (roughly  the recoiling $u$ quark) direction. 
 At this order,  effects such as  the ``jiggling'' of  $k_\perp$ are ignored and the
 differential partial width is given by the convolution of the shape function with the
 parton level differential distribution.
Because the shape function depends only on parameters of the $B$ meson, this
leading order description holds for any $B$ decay to a light quark. It holds, in particular, 
for $B\to s\gamma$, which can provides an estimation of $f(k_+)$ via
the shape of the photon energy ($E_\gamma$) 
spectrum\cite{bigiEndpoint}\cite{neubertEndpoint}.  In addition to the increased uncertainty
on $\Vub$ from the $m_b$ and $b$ quark kinetic energy contributions that results from the 
restriction of phase space, higher twist contributions and unknown power corrections of order 
$\Lambda_{QCD}/M_B$\cite{Leibovich:2000kv}\cite{neubertNote} also contribute  to the 
uncertainty, as will be discussed further below.
%
%

Ideally, $\Vub$ would be
determined without introduction of an intermediate extracted shape function
through the use of appropriately weighted spectra\cite{neubertEndpoint}\refrange{bb:rothstein_endpoint}{bb:bigi_weighted_integrals}.  This would avoid
introduction of an element of model dependence.
For the lepton spectrum, for example, one would take
$$
\left| \frac{V_{ub}}{V_{tb} V_{ts}^*} \right|^2 
   = \frac{3\alpha}{\pi}\,K_{\rm pert}
   \frac{\widehat\Gamma_u(E_0)}{\widehat\Gamma_s(E_0)}
   + O(\Lambda_{\rm QCD}/M_B),
\EQN{eq:integrate}
$$
where $K_{{\rm pert}}$ is a calculable perturbative kernel, abd $\widehat\Gamma_u(E_0)$ and $\widehat\Gamma_s(E_0)$ are appropriately
weighted integrals over, respectively,  the lepton energy and photon energy spectra
above the minimum cutoff energy $E_0$.  
 Practical application of this approach awaits measurement of the lepton momentum
 spectrum in the $B$, not the $\Upsilon(4S)$, rest frame, which $B$ tag methods
 will permit in the future. Similar expressions exist for integration over the low hadronic
 mass region\cite{bb:Rothstein_MX}\cite{bb:bigi_weighted_integrals}, so, in principle, 
 current $M_X$ analyses should be able to take such an approach. Experimental efficiency and
 lepton momentum cutoffs must, however, be incorporated into the integrals.  To date,
experiments have instead introduced intermediate shape functions of a variety of forms
into the inclusive analyses, as discussed below.
 
A third way to isolate the charmless s.l.\ signal is to use a selection based on the $q^2$ of 
the leptonic system.
Restriction of phase space to regions of large $q^2$ also restores the validity of the 
OPE\cite{bb:hepph0107074}\cite{bb:bll1} and suppresses
shape function effects.  Taking only the region kinematically forbidden to $\btoclnu$,
$q^2>(M_B-M_D)^2$, unfortunately introduces a low mass 
scale\cite{bb:original_neubert_mccubed}\cite{bb:neubert_mccubed}
into the OPE and the $1/m^3$ uncertainties blow up to be of order $(\Lambda_{QCD}/m_c)^3$.
However, a combination of $M_X$ with looser $q^2$ requirements can suppress both
$\btoclnu$ background experimentally and shape function effects theoretically.  Furthermore,
the $q^2$ requirement moves the parton level pole away from the experimentally
feasible $M_X$ requirement.   
%
The shape function effects, while suppressed, cannot be neglected.  One drawback
of the $q^2$ requirement is the elimination of higher energy hadronic
final states, which may exacerbate duality concerns.

A recent BELLE analysis\cite{bb:belle_mxq2} has been performed in this region.  BELLE
employs a $p_\ell>1.2$ GeV$/c$ requirement in the $\Upsilon(4S)$ rest frame,
 and an ``annealing'' procedure to separate reconstructed particles into signal 
 and ``other B'' halves.   They then examine the integrated rate in the region
 $M_X<1.7$ GeV and $q^2>8$ GeV$^2$ to extract $\Vub$, which again has the
 desired effect of minimizing dependence of the analysis on detailed $\btoulnu$ modeling
 The signal to background ratio of the annealing technique,
about 1:6 for the BELLE analysis, is significantly degraded relative to that of the 
hadronic $B$ tag technique.   As we mentioned in the previous review, control  of
background subtractions of this size requires extreme care and careful scrutiny
of the associated systematic issues.
BELLE finds the rate $\Delta{\cal B}$ in that region of phase space to be
$$
\Delta{\cal B} = (7.37\pm0.89_{\rm stat} \pm 1.12_{\rm sys} \pm0.55_{c\ell\nu}
\pm0.24_{u\ell\nu})\times 10^{-4}.
$$
An analysis of this restricted region of phase space, for which the shape function
influence is significantly reduced\cite{bb:hepph0107074}, with the significantly cleaner $B$
tag technique should be a priority for both $B$ experiments.

Each analysis discussed here has relied on an intermediate shape function to evaluate
the fraction of the inclusive rate that lies in its restricted region of phase space.
 The endpoint analyses have used rate fractions  \Ref{bb:cleo_endpoint}  based
 on intermediate shape functions derived from the CLEO $b\to s\gamma$ photon
 spectrum. Several two-parameter ansaetze\cite{bigi94}\cite{kagan98},
 $F[\Lambda^{SF},\lambda_1^{SF}]$, were implemented as
 the form of the shape function.  
These parameters are related to the HQET parameters of similar name, and play a similar role in
evaluation of the rates.  At this time, however, we do not know the precise relationship
between the shape function parameters, or the moments of the
shape function, and the HQET nonperturbative parameters $\overline{\Lambda}$ and
$\lambda_1$\cite{bb:aMatthiasReference}\cite{bb:hepph0312109}.  The fact
that $\Lambda^{SF}$ and $\lambda_1^{SF}$ depend on the functional ansatz
while  the HQET parameters depend on the renormalization scheme underscores
the current ambiguity. 

With the limited $E_\gamma$
statistics available, there exists a strong correlation between the two parameters
because of  the interplay between the effective $b$ quark mass (controlled by $\Lambda^{SF}$)
and the effective $b$ quark kinetic energy (controlled by $\lambda_1^{SF}$) in determining
the {\it mean} of the $E_\gamma$ spectrum.  No external constraints that could break
the correlation, such as $m_b^{(1S)}$ or measured $\btoclnu$ moments, have 
been input into the $b\to s\gamma$ fits because of their unknown relationship to the 
shape function parameters.  The resulting effective $m_b$ mass range contributing to the 
uncertainties ($\pm200$ MeV) is therefore much larger than the current $m_b$ uncertainty.
Given the current independence of the shape function and $m_b$ determinations and
the broad effective $m_b$ range sampled in the shape function, we do not consider it 
necessary to treat the $E_\gamma$--derived phase space fractions and the
partial width (\Eq{eq:rate}) as positively correlated\cite{bb:belle_mxq2}.
As the data statistics increase, it will become possible to constrain the shape function 
parameters directly from distributions in $\btoulnu$, such as the $M_X$ spectrum, thus removing 
the uncertainties introduced from their derivation from another class of decays.
Furthermore, once the renormalization behaviour of the shape function and the relationship
of its moments to the HQET parameters becomes known, the powerful constraints
from the kinematic mass of the $b$ quark and from moments information in the
$b\to c\ell\nu$ system can be incorporated into a shape function derivation
based either on $b\to s\gamma$ or $\btoulnu$.


Two alternate approaches to the shape function evaluation have been taken in experimental
studies so far. In their low-$M_X$ analysis\cite{bb:babar_mx}, BaBar has evaluated the phase space
fraction using the same $f(k_+)$ parameterizations noted above, but has substituted HQET
parameters derived from studies of spectral moments of the $b\to s\gamma$ and
$\btoclnu$ processes.
The BELLE  $M_X$--$q^2$ analysis\cite{bb:belle_mxq2} (discussed below)
uses the calculation of Bauer \etal\cite{bb:hepph0107074} based on a form
with a single parameter $a=\Lambda^{SF}/\lambda_1^{SF}$, which was estimated from
the $m_b^{(1S)}$ mass and from typical estimates for $\lambda_1$.
The uncertainties in the different rate fractions in the momentum endpoint, 
the low $M_X$ and the $M_X$--$q^2$ regions are strongly correlated, and
the values and uncertainties are sensitive to the theoretical assumptions made.
Hence, a common theoretical scheme must be
chosen for meaningful comparison of the extracted values of $\Vub$.
Given the {\it ad hoc} nature of the association of shape function parameters
with the HQET parameters in the $\overline{MS}$ or the
$\Upsilon(1S)$ mass scheme\cite{Hoang:1998ng}, and the difficulty in evaluating the
uncertainty in such an association, we have chosen to extract $\Vub$ from all of the 
measurements discussed using the  $b\to s\gamma$--derived shape function.    

\midtable{tab:inclusivesummary}
\Caption
Summary of inclusive $\Vub$ measurements.  The last five measurements
are incorporated into the analysis presented below.  The errors in the first group
are the experimental and theoretical uncertainties. The errors in the second group
are from the statistical, experimental systematic, $E_\gamma$--based rate 
fraction, and $\Gamma_{\rm tot}$  uncertainties. The two groups are {\it not} directly comparable
as they have not been evaluated with identical theoretical inputs.
\endCaption
\vglue -18pt
\centerline{
\vbox{
\offinterlineskip
\halign{\strut#&\tabskip=.5em plus 1em minus 0.5em
\centertab{#}& \lefttab{#} \tabskip=0pt\cr
\tableheaddoublerule
& $\Vub (10^{-3})$ & \cr
\tableheadsinglerule
ALEPH~[53]  & $4.12 \pm 0.67 \pm 0.76$ & neural net \cr
L3~[54]           & $5.70 \pm 1.00 \pm 1.40$ & cut and count \cr
DELPHI & $4.07 \pm 0.65 \pm 0.61$ & $M_X$  \cr
OPAL~[55]     & $4.00 \pm 0.71 \pm 0.71$ & neural net \cr
LEP Avg. &  $4.09 \pm 0.37 \pm 0.56$ & \cr
CLEO~[56]     & $4.05 \pm 0.61 \pm 0.65$ & $d\Gamma/dq^2dM_X^2dE_\ell$ \cr
BELLE   & $5.00 \pm 0.64 \pm 0.53$ &  $M_{X}$, $D^{(*)}\ell\nu$ tag \cr
CLEO     & $4.11 \pm 0.13 \pm 0.31 \pm 0.46 \pm 0.28$ &  $2.2 < p < 2.6$ \cr
BaBar  & $4.31 \pm 0.20 \pm 0.20 \pm 0.49 \pm 0.30$ & $2.3 < p < 2.6$  \cr
BELLE   & $3.99 \pm 0.17 \pm 0.16 \pm 0.45 \pm 0.27$ & $2.3 < p < 2.6$  \cr
BELLE   & $4.63 \pm 0.28 \pm 0.39 \pm 0.48 \pm 0.32$ & $M_X<1.7$, $q^2>8$ \cr
BaBar  & $4.79 \pm 0.29 \pm 0.28 \pm 0.60 \pm 0.33$ & $M_{X}<1.55$ \cr
\tableheaddoublerule
} 
} 
} 
\endtable

The full set of inclusive $\Vub$ results is summarized in  \Tbl{tab:inclusivesummary}, 
which is an updated version of the Heavy Flavors Averaging Group
summary\cite{bb:hfag_vub} .  All endpoint results have  QED radiative corrections applied correctly.
The listed uncertainties  do not include contributions for potentially large
theoretical corrections that have been categorized but remain incalculable (see below).  
The last five results in the table, which we
will use below, have been updated to a common framework
based on the CLEO $E_\gamma$--derived shape function.  The
rate fractions\cite{bb:gibbonsbeauty} for the  BaBar $M_X$ analysis ($f_M$)
and the BELLE $M_X-q^2$ analysis $f_{qM}$  are $f_M=0.55\pm0.14$ and $f_{qM}=0.33\pm0.07$.  
The central values for these and for the endpoint fractions (\Tab{tab:endpointRates})
correspond to an exponential shape function
ansatz\cite{kagan98} and $(\lambda_1^{SF},\bar\Lambda^{SF})=(-0.342,0.545)$, with
small corrections related to background subtractions in the $b\to s\gamma$ spectrum.
The errors are dominated by the statistical uncertainty in the $f(k_+)$
fit to the $E_\gamma$ spectrum, but include contributions from experimental systematics,
$\alpha_s$ uncertainties and modeling. Incorporation
of results beyond those used here will require significant input from the 
experimental analyses, and is left to the HFAG.

\subsection{Combining inclusive information}

Evaluation of the total uncertainty on $\Vub$ remains problematic because of
a variety of theoretical complications.  A recent review\cite{bb:luke_ckm}
discusses these issues in detail.  There are three main contributions.  The
first arises from subleading (higher twist) contributions to the shape function
resummation \refrange{LLW_SSF}{BLT_SSF1}. 
These involve incorporation of effects 
such as the variation of $k_\perp$,
and are not universal for all $B$ decay processes.  Hence with the
use of $b\to s\gamma$ to obtain a shape function, there are two contributions, one
from subleading contributions to the use of a shape function in $\btoulnu$ process itself, 
and the second from the different corrections in $b\to s\gamma$ from which the
shape function is obtained.  These contributions are potentially large, since they
are of order $\Lambda_{QCD}/m_b$.  Indeed, a partial estimate of
these effects\cite{N_SSF} for the momentum endpoint region finds corrections that
are similar in size to the total uncertainties of those analyses.

The second contribution, from ``weak annihilation'' processes, is
formally of order  $(\Lambda_{QCD}/m_b)^3$ but receives a large multiplicative 
enhancement of $16\pi^2$\cite{bigi_wa}\cite{voloshin_wa}.
The contribution, which requires factorization violation to be nonzero, is expected to be 
localized near $q^2\sim m_b^2$, and this localization
can result in a further enhancement of the effect on $\Vub$.  For
the endpoint region, which sees about 10\% of the total rate, an effect on the total
rate of 2-3\% (corresponding to factorization violation of about 10\%), produces an effect on
the measured rate of 20-30\%. 

Finally, there are unknown contributions from potential violation of local
quark hadron duality.  The true differential distribution cannot be predicted
via the OPE -- the resonant substructure is not described.  However, spectra
integrated over a sufficiently broad range should be better described.

The problems just outlined present a challenge to the averaging of the  
various inclusive results.   Results with a potentially large bias might be 
included with neither a correction nor an appropriate uncertainty due to these effects.  
The resulting $\Vub$ determination would be potentially biased and the attached 
uncertainty unreliable. This motivated us not to provide an average result in the 
first edition of this review two years ago.

As an alternative, we here choose measurements in the region of phase space that
appears to have the best compromise of the affects discussed to
to obtain an estimate of $\Vub$.  Measurements from the other regions of phase space, which 
have increased  sensitivity to one or more of the corrections, then provide 
limits on the uncertainties from these effects and thereby allow
as complete as possible an estimation of
the theoretical uncertainty, as first proposed by~\Ref{bb:gibbonsbeauty}.
At this time, the low $M_X$, high $q^2$ region appears to be the best 
motivated choice. It has  reduced (though by no means negligible) corrections from the shape function and thus also from
the subleading contributions to the shape function. Yet it integrates over a sufficient fraction 
of the spectrum to dilute weak annihilation contributions and concerns on local quark hadron duality.

While this choice is at present subjective, it
offers the advantage of a reduction of the shape
function influence coupled with the ability to bound
the remaining theoretical uncertainties.
In the opinion of the reviewers, this is a
reasonable tradeoff for the statistical loss
relative to the low $M_X$ region. We expect that
each experiment will perform an improved combination of
information from the different regions of phase
space where the experimental and theoretical
correlations can be made manifest more
straightforwardly.

We further stress that we view all three regions as equally crucial in this
combination of information, as a more complete evaluation of the inclusive
uncertainty than has previously existed is necessary for proper use of
the inclusive results.  The choice of the phase space region should not
be misconstrued as a preference of experimental technique.  Indeed,
we look forward to a similar (or improved) analysis when a sample
of clean results based on  fully tagged $B$ samples have been
obtained for all regions of phase space.

At present only BELLE\cite{bb:belle_mxq2} has contributed a result for 
this region of phase space,  so for now we take this result as the ``central value'':
$$
\eqalign{
\Vub/10^{-3} &= 4.63 \pm 0.28_{\rm stat} \pm 0.39_{\rm sys} \rm 0.48_{f_{qM}}
 \pm 0.32_{\Gamma{\rm thy}} \cr
& \pm \sigma_{\rm WA} \pm \sigma_{\rm SSF} \pm \sigma_{\rm LQD}.
}
$$
Additional measurements by the $B$ factories of the rate in this region of phase space
will soon improve the experimental uncertainties. 

We must determine the last three uncertainties for weak annihilation (WA), subleading
shape function corrections (SSF) and local quark hadron duality (LQD).  
The measurements from other regions of phase space are crucial for this task.

We assume that the WA contribution is largely contained within
each of the $p_\ell> 2.2$ GeV/$c$, the $M_X<1.55$ GeV and the combined
$M_X<1.7$ GeV, $q^2>8$ GeV$^2$ regions.   The contribution will be most
diluted in the low $M_X$ region, with the rate fraction $f_M=0.55\pm0.14$, and most
concentrated in the endpoint region, with the rate fraction $f_e=0.14\pm0.03$
(without radiative corrections).  It is
simple to show that for a neglected WA contribution, a
comparison of $\Vub$ from these two regions would predict the bias in
the $M_X$, $q^2$ region (with rate fraction $f_{qM}=0.33\pm0.07$) to be
$$
[(1-f_{qM})/f_{qM}][f_ef_M/(f_M-f_e)] \approx 0.39
\EQN{eq:wa_scale}
$$
of the observed difference.
Comparison of the endpoint result from CLEO and the low $M_X$ result from 
BaBar, taking into consideration
the almost total correlation in the shape function and $\Gamma_{tot}$ uncertainties,
yields $\Delta|V_{ub}|/10^{-3} = 0.69 \pm 0.53$.  There is not sufficient sensitivity
to draw conclusions regarding the presence of a WA component, but we can
place a bound.
We take the larger of the error and central value and scale according to \Eq{eq:wa_scale} 
to obtain
$$
\sigma_{\rm WA} \approx 0.27.
$$

To estimate the uncertainty from the subleading corrections to the shape function, we
assume that subleading corrections will scale like the fractional change in the
predicted rate ($\Delta \Gamma/\Gamma$) with and without  convolution of the 
parton--level expression with the shape function.
As the base comparison, we take the low $M_X$ region,
with $(\Delta \Gamma/\Gamma)_M=0.15$, and compare to the combined $M_X$, $q^2$
region, with $(\Delta \Gamma/\Gamma)_{qM}=-0.075)$.  The shifts again depend on the
shape function modeling, and the quoted values correspond to the $f(k_+)$ from the best fit 
to the CLEO $E_\gamma$ spectrum . The theory uncertainties are
again correlated, and we find $\Delta|V_{ub}|/10^{-3} = 0.16 \pm 0.63$.   Scaling the uncertainty
of the comparison by $|(\Delta \Gamma/\Gamma)_{qM}/(\Delta \Gamma/\Gamma)_M|=0.49$,
we have
$$
\sigma_{\rm SSF} \approx 0.31.
$$

Finally, we must make an estimate of the local duality uncertainty.  We
assume that  a potential violation will scale with the fraction of rate $f$ in a given region
as $(1-f)/f$.  This form ranges from no ``local'' violation for integration of the full phase space
($f=1$), to large uncertainty for use of a very localized region of phase space ($f\to 0$).
The estimate derives from comparison of
 the CLEO $p_\ell>2.2$ GeV/$c$ analysis
($f\sim0.14\pm0.03$)  to the average of the BaBar and BELLE $p_\ell>2.3$ GeV/$c$
analyses ($f\sim0.07\pm0.02$).  The subleading correction 
estimates of \Ref{N_SSF} are applied to minimize potential cancelation between
duality violation and subleading corrections.  This yields $(\Vub^{2.3}-\Vub^{2.2}+0.27)/10^{-3}=0.29\pm0.38$, where the 0.27 is the estimate of the relative subleading
correction.  With our scaling assumption, we then apply a scale factor $s$ of
$$
s = \frac{(1-f_{qM})/f_{qM}}{(1-f_{2.3})/f_{2.3}-(1-f_{2.2})/f_{2.2}} \approx 0.29
$$
to the uncertainty in this difference estimate.  Our local duality estimate therefore is
$$
\sigma_{\rm LQD} \sim 0.11.
$$

From this analysis, we finally obtain
$$
\eqalign{
\Vub/10^{-3} &= 4.63 \pm 0.28_{\rm stat} \pm 0.39_{\rm sys} \pm 0.48_{f_{qM}}
 \pm 0.32_{\Gamma{\rm thy}} \cr
& \quad \pm 0.27_{\rm WA}  \pm 0.31_{\rm SSF}  \pm 0.11_{\rm LQD},
}
$$
for a total theory error of 15\% and total precision of 18\%.  
Given that the uncertainties are dominated by
experimental limits, addition in quadrature seems appropriate.   Note that these
estimates apply {\it only} in the combined low $M_X$, high $q^2$ region of phase space.
The limits presented here can be improved both in robustness, through more sophisticated 
scaling estimates, and in magnitude, through additional and improved $\Vub$ measurements and
through inputs from other sources.  The consistency of the values of $\Vub$ extracted 
with different inclusive methods and the stability of the results over changes in the 
selected region of phase space will provide increasing confidence in the reliability
of the extracted results and of their estimated uncertainties. Improvement of the $b\to s\gamma$ 
photon energy spectrum is key until a self-consistent extraction of the shape function from 
$\btoulnu$ transition becomes available.  Comparisons of the
$D^0$ versus $D_s$ semileptonic widths and of the rates for charged versus
neutral $B$ mesons  can provide estimates of the weak 
annihilation contributions\cite{voloshin_wa}.
Finally, improved theoretical guidance concerning the
scaling of the effects over phase space would allow development of a simultaneous
extraction of $\Vub$ and the corrections, with all experimental information contributing 
directly to $\Vub$.

\section{Exclusive measurements of $\btoulnu$}

Reconstruction of exclusive $b\to u\ell\bar{\nu}$ channels provides
powerful kinematic constraints for suppression of the $b\to c\ell\bar{\nu}$
background.  For this suppression to be effective, an estimate of 
the four momentum of the undetected neutrino must be provided.
The measurements to date have made use of detector hermeticity
and the well-determined beam parameters to define a missing
momentum that is used as the neutrino momentum. Signal-to-background ratios 
(S/B) of order two have been achieved in these channels.

To extract $\Vub$ from an exclusive channel, the
form factors for that channel must be known.  The form factor normalization
 dominates the uncertainty on $\Vub$.  The $q^2$--dependence of the form factors,
which is needed to determine the experimental efficiency, also
contributes to the uncertainty, but at a much reduced level.  For example, the
requirement of a stiff lepton for background reduction in these analyses introduces a
$q^2$--dependence to the efficiency. 
In the limit of a massless charged lepton (a reasonable limit for the
electron and muon decay channels), the $B\to\pi\ell\nu$ decay depends on
one form factor $f_1(q^2)$:
$$
\frac{d\Gamma(B^0\to\pi^-\ell^+\nu)}{dy\,d\cos\theta_\ell} =
\Vub^2\frac{G_F^2 p_\pi^3 M_B^2}{32\pi^3}\sin^2\theta_\ell|f_1(q^2)|^2,
$$
where $y = q^2/M_B^2$ and $\theta_\ell$ is the angle between the charged lepton
direction in the virtual $W$ ($\ell+\nu$) rest frame and the direction of the
virtual $W$.  For the vector meson final states $\rho$ and $\omega$,
three form factors $A_1$, $A_2$ and $V$ are necessary (see {\it e.g.}
reference\cite{Gilman:1989uy}).

Calculation of these form factors constitutes a considerable
theoretical industry, with a variety of techniques now being employed.
Form factors based on lattice QCD calculations\refrange{Abada:1993dh}{Aoki:2001rd}
and on light cone sum rules\refrange{Ball:1997rj}{Ball:2001fp}
currently have uncertainties in the $15\%$ to $20\%$ range.  A variety
of quark model calculations exists\refrange{Wirbel:1985ji}{Choi:1999nu}. Finally,
a number of other approaches\refrange{Kurimoto:2001zj}{Mannel:1998kp}, such as 
dispersive bounds and experimentally--constrained models based on Heavy Quark Symmetry,
seek to improve the $q^2$ range where the form factors can be estimated, without
introducing significant model dependence. 

Of particular interest are the light cone sum rules (LCSR) and 
lattice QCD (LQCD) calculations, which minimize
modeling assumptions as they are QCD--based calculations and provide
a much firmer basis compared to the quark model calculations for systematic
evaluation of the uncertainties.  The calculations used in the current results
have been summarized nicely in \Ref{Battaglia:2003in}. The LCSR
are expected to be valid in the region $q^2\simle 16$ GeV$^2$.
The light cone sum rules calculations use quark--hadron duality to estimate some
spectral densities, and offer
a ``canonical'' contribution to the related uncertainty of $10\%$ with no known
means of rigorously limiting that uncertainty.   The theory community is currently
debating the size of potential contributions to the form factors missing from
the LCSR approach\refrange{bb:ph0310139}{bb:ph0311345} that have been revealed using the newly--developed
soft collinear effective theory 
(SCET).  The
$B\to\rho\ell\nu$ form factors, in particular, could be appreciably overestimated,
biasing $\Vub$ low.  Two exclusive results will therefore be presented in this
review, one based on the full set of exclusive results, and the second based only
on results in the $q^2>16$ GeV$^2$ region for $B\to\rho\ell\nu$.

The LQCD calculations that can be applied to experimental $B\to X_u\ell\nu$ decay
remain, to date,  in the ``quenched'' approximation (no light quark loops in the 
propagators), which limits the ultimate precision to the 15\% to 20\% range.  The
$q^2$ range accessible to these calculations has been $q^2\simge 16$ GeV$^2$
Significant progress has been made towards unquenched lattice QCD calculations,
and a recent comparison\cite{bb:lat0304004} of a range unquenched results to 
experiment shows much better agreement (few percent) than the corresponding
quenched results.  Work has begun on the unquenched form factors needed
for $\Vub$, though the initial results have been limited to valence quarks
closer to the strange quark mass.  Nevertheless, initial results 
\cite{bb:lat0309055}\cite{bb:lat0309107} are compatible with the 15\% to 20\% uncertainties
used for the quenching uncertainty, lending them some validity.

The exclusive $\Vub$ results are summarized in \Tbl{tab:exclusiveVub}.
These include a simultaneous measurement of the
$B \to \pi \ell \bar \nu$ and the $B \to \rho \ell \bar \nu$
transitions by CLEO\cite{cleo-pilnu}, and measurement of the
$B \to \rho \ell \bar \nu$ rate by CLEO\cite{cleo-rholnu} and BaBar\cite{babar-rholnu}. 
All measurements
employ the missing energy and momentum to estimate the neutrino momentum.
With that technique, the major background results from $b\to c\ell \bar{\nu}$ decays
in events that cannot be properly reconstructed (for example, because of
additional neutrinos in the event) and hence which overestimate the
neutrino energy. All measurements also employ the isospin relations
$$
\Gamma(B^0\to\pi^-\ell^+\nu) = 2\Gamma(B^+\to\pi^0\ell^+\nu) 
$$
and
$$
\Gamma(B^0\to\rho^-\ell^+\nu) = 2\Gamma(B^+\to\rho^0\ell^+\nu) 
$$
to combine the charged and neutral decays.  These relationships can
be distorted by $\rho-\omega$ mixing\cite{bb:DiazCruz}, and all results
discussed here allow for this possibility in their systematic evaluation.

{
\midtable{tab:exclusiveVub}
\Caption
Summary of all exclusive $\Vub$ measurements.  
For the CLEO '00 and BaBar '01 measurements, the errors arise from
 statistical, experimental systematic  and form factor modeling uncertainties, respectively.
For the CLEO '03 measurements, the errors arise from statistical, experimental systematic,  $\rho\ell\nu$ form factor, and LQCD and LCSR calculation uncertainties, respectively.
In the CLEO '03 averages, the LQCD and LCSR uncertainties have been treated as correlated.
\endCaption
\vglue -18pt
\centerline{
\vbox{
\offinterlineskip
\halign{\strut#&\tabskip=.5em plus 1em minus 0.5em
\lefttab{#}& \centertab{#}&  \centertab{#}& \lefttab{#} \tabskip=0pt\cr
\tableheaddoublerule
& mode & $\Vub (10^{-3})$ & $q^2$ range & FF \cr 
\tableheadsinglerule
CLEO '00 & $\rho\ell\nu$ & $3.23\pm0.24^{+0.23}_{-0.26}\pm0.58$ & all & model survey \cr
BaBar '01 & $\rho\ell\nu$ & $3.64\pm0.22\pm 0.25^{+0.39}_{-0.56}$ & all & model survey \cr
CLEO '03 & $\pi\ell\nu$ & $3.33\pm 0.24 \pm0.15\pm 0.06 \;^{+0.57}_{-0.40}$ & $q^2<16$ GeV$^2$ & LCSR \cr 
CLEO '03 & $\pi\ell\nu$ & $2.88\pm 0.55\pm 0.30\pm 0.18\;^{+0.45}_{-0.35}$ & $q^2>16$ GeV$^2$ & LQCD \cr
CLEO '03 & $\pi\ell\nu$ & $3.24\pm0.22\pm 0.13\pm0.09 ^{+0.55}_{-0.39}$ & {average} \cr
CLEO '03 & $\rho\ell\nu$ & $2.67\pm 0.27\;^{+0.38}_{-0.42} \pm 0.17 \;^{+0.47}_{-0.35}$ & $q^2<16$ GeV$^2$ & LCSR \cr 
CLEO '03 & $\rho\ell\nu$ & $3.34\pm 0.32\;^{+0.27}_{-0.36}  \pm 0.47 \;^{+0.50}_{-0.40}$ & $q^2>16$ GeV$^2$ & LQCD \cr
CLEO '03 & $\rho\ell\nu$ & $3.00\pm0.21^{+0.29}_{-0.35}\pm0.28 ^{+0.49}_{-0.38}$ & {average} \cr
CLEO '03 & $\pi+\rho$ & $3.17\pm0.17^{+0.16}_{-0.17}\pm0.03 ^{+0.53}_{-0.39}$ & {average} \cr
CLEO '03 & $\pi+\rho$ & $3.26\pm0.19\pm0.15\pm0.04 ^{+0.54}_{-0.39}$ & average & no $\rho\ell\nu$ LCSR \cr
\tableheaddoublerule
} 
} 
} 
\endtable
}

In the combined $\pi$ and $\rho$ measurement,
strict event quality requirements were made that resulted in a low
efficiency, but a relatively low background to signal ratio over a fairly broad
lepton momentum range.  
The $\rho$-only analyses employ looser event cleanliness requirements,
resulting in a much higher efficiency.  The efficiency gain comes at
the price of an increased background, and the analyses are primarily
sensitive to signal with lepton momenta above 2.3 GeV$/c$, which is
near (and beyond) the kinematic endpoint for $b\to c\ell\bar{\nu}$ decays which are 
therefore highly suppressed.  

The combined $\pi$ and $\rho$ analysis of CLEO employs relatively loose lepton
selection criteria and extracts rates independently in three separate $q^2$ intervals.
Form factor dependence of the rates is then evaluated using models and calculations that
exhibit a broad variation in $d\Gamma/dq^2$, which shows that  this approach has
eliminated model dependence of the rates in $\pi\ell\nu$, and significantly reduced it
in $\rho\ell\nu$.    To further reduce modeling uncertainties, CLEO then extracts $\Vub$
using only the LQCD and LCSR QCD-based calculations  restricted to their respective
valid $q^2$ ranges, thereby eliminating modeling used for extrapolation.  Averages
of the CLEO results, with and without the low $q^2$ region for $\rho\ell\nu$ are
listed in \Tbl{tab:exclusiveVub}.

A more complete review of recent $B\to X_u\ell\nu$ branching fractions,
including analyses too incomplete for inclusion in this $\Vub$ summary,
can be found in Reference\cite{bb:gibbons_ckm}.  Of note is the recent
evidence presented for $B\to\omega\ell\nu$ by BELLE\cite{bb:belle_omega}.

With all results resting on use of detector hermeticity, the potential for significant
correlation among the dominant experimental systematics exists\cite{bb:gibbons_ckm}.
Results from the three measurements have been averaged here assuming full
correlation in these systematics.   The $\rho\ell\nu$-only 
results\cite{cleo-rholnu}\cite{babar-rholnu},
 which depend more heavily on modeling even for the LCSR and LQCD calculations,
 are deweighted by 5\% in the average.  This yields
$$
|V_{ub}| = (3.27 \pm 0.13 \pm0.19 ^{+0.51}_{-0.45}) \times 10^{-3}
$$
where
the  errors arise from statistical, experimental systematic  and form factor
uncertainties, respectively.  While similar in precision to the exclusive result in
the previous $\Vub$ minireview, this result relies much less heavily on modeling.
Should the LCSR form factors prove to be overestimated, we also provide an
average excluding any result using information for $q^2<16$ GeV$^2$ in the
$\rho\ell\nu$ modes, with the result
$$
|V_{ub}| = (3.26\pm0.19\pm0.15\pm0.04 ^{+0.54}_{-0.39}) \times 10^{-3},
$$
where
the  errors arise from statistical, experimental systematic  $\rho\ell\nu$ form factor
uncertainties, and LQCD and LCSR (treated as correlated), respectively. 

The future for exclusive determinations of $\Vub$ appears promising. Unquenched lattice 
calculations are appearing, with very encouraging results. These calculations will 
eliminate the primary source of uncontrolled uncertainty in these calculations, and have
already provided some validity to the quenching uncertainty estimate used in the results
presented here. Simultaneously, the $B$ factories
are performing very well, and very large samples of events in which one $B$ meson
has been fully reconstructed are already being used.  This will allow a more robust
determination of the neutrino momentum, and should allow a significant reduction of 
backgrounds and experimental systematic uncertainties.  The high statistics should
also allow more detailed measurements of $d\Gamma/dq^2$, which have already provided
a sorely-needed
litmus test for the form factor calculations and reduced the form factor shape contribution
to the uncertainty on $\Vub$.  Should theory allow use of the full range of
$q^2$ in the extraction of $\Vub$\cite{bb:lepage_moving_B}, the $B$ factories have already logged data sufficient for a 5\% statistical determination of $\Vub$. 

For both lattice and the $B$ 
factories, $\pi\ell\nu$ appears to be a golden mode for future precise determination of
$\Vub$. The one caveat is management of contributions from the $B^*$ pole, but recent
work\cite{El-Khadra:2001rv} suggests that this problem can be successfully overcome.
$B\to \eta\ell\nu$ will provide a valuable cross-check.
The $\rho\ell\nu$ mode will be more problematic for high precision: the broad width 
introduces both experimental and theoretical difficulties.  
Experiments must, for example, 
assess potential nonresonant $\pi\pi$ contributions, but only crude arguments based on
isospin and quark-popping have been brought to bear to date.  Theoretically,
no calculation, including lattice, has dealt with the width of the $\rho$.  When the lattice calculations become unquenched, the $\rho$ will become unstable
and the $\pi\pi$ final state must be faced by the calculations.  The methodology
for accommodation of high-energy two particle final states on the lattice has yet to be developed.
The $\omega\ell\nu$ mode
may provide a more tractable alternative to the $\rho$ 
mode because of the relative narrowness of the $\omega$ resonance. Agreement between accurate $\Vub$ 
determinations from $\pi\ell\nu$ and from $\omega\ell\nu$ will provide added confidence 
in both.

\section{Combined results}

The experimental bounds provided for the outstanding uncertainties in the inclusive
$\Vub$ measurements in the low $M_X$, high $q^2$ region and theoretical work which
clarifies the reliability of the LCSR and quenched LQCD form factors make possible 
comparison of the inclusive and exclusive determinations of $\Vub$. Results agree to better 
than 1.5 times the quadratic combination of the quoted uncertainties. Therefore it becomes
feasible to propose an average the inclusive and exclusive results, which have comparable accuracies. 
 
The uncertainties have been combined in quadrature, using the larger (upward) error for the 
exclusive numbers.  The proposed average of the inclusive and exclusive results, with all the 
exclusive data considered, is
$$
\Vub = (3.67\pm0.47)\times 10^{-3}.
$$
Including in the average only the exclusive analyses based on data with
$q^2>16$ GeV$^2$ in the  $\rho\ell\nu$ mode, the average becomes
$\Vub = (3.70\pm0.49)\times 10^{-3}$,
so exclusion of this region, if appropriate, has only a minor effect.

The procedure proposed here results in a value of $\Vub$ with a 13\% uncertainty.
With the experimental and theoretical progress expected over the next few years 
an improvement of the accuracy at the 10\% level, and possibly below, appears now 
realizable.  

\section{References}

\ListReferences
\end